\documentclass[a4paper,pra, twocolumn,superscriptaddress, nofootinbib,showpacs,showkeys,amsfonts,amssymb,amsmath]{revtex4-2}
\usepackage{amssymb,latexsym}

\usepackage{color}
\usepackage{amscd}
\usepackage{times}
\usepackage{epsfig}
\usepackage{psfrag}
\usepackage{graphicx}
\usepackage{amsthm,amsmath}
\usepackage{mathrsfs}
\usepackage[colorlinks=true, allcolors=blue]{hyperref}
\usepackage{leftindex}
\usepackage{mathtools}
\usepackage[dvipsnames]{xcolor}
\usepackage{physics}
\usepackage{hyperref}
\usepackage{units}
\usepackage{csquotes}

\begin{document}

\title[]{White paper: 1-10 Hz matter-wave interferometer to test the spin entanglement witness for quantum gravity}

\author{Sougato Bose}
\affiliation{Department of Physics and Astronomy, University College London, London WC1E 6BT, United Kingdom}
\affiliation{David Potter Institute for Quantum Information and Spacetime, University College London, Gower Street, WC1E 6BT London, United Kingdom}

\author{Anupam Mazumdar}
\email{mazumdar@cita.utoronto.ca}
\affiliation{Canadian Institute for Theoretical Astrophysics, 60 St George St, University of Toronto, Toronto, ON M5S 3H8, Canada}

\author{Marko Toro\v{s}}
\affiliation{Faculty of Mathematics and Physics, University of Ljubljana, Jadranska 19, SI-1000 Ljubljana, Slovenia}

\author{Tian Zhou}
\affiliation{Van Swinderen Institute, University of Groningen, Nijenborg 3, 9747 AG Groningen, The Netherlands}

\author{Tadeusz Adach}
\affiliation{University of Wroc\l{}aw, Faculty of Physics and Astronomy, pl.\ M.\ Borna 9, 50-204 Wroc\l{}aw, Poland}

\author{Niayesh Afshordi}
\affiliation{Department of Physics and Astronomy, University of Waterloo, 200 University Avenue West, Waterloo, ON N2L 3G1, Canada}
\affiliation{Waterloo Centre for Astrophysics, University of Waterloo, Waterloo, ON N2L 3G1, Canada}
\affiliation{Perimeter Institute for Theoretical Physics, 31 Caroline St. North, Waterloo, ON N2L 2Y5, Canada}

\author{Agya Sewara Alam}
\affiliation{Van Swinderen Institute, University of Groningen, Nijenborg 3, 9747 AG Groningen, The Netherlands}

\author{Alexandre Arbey}
\affiliation{Institut de Physique des 2 Infinis de Lyon (IP2I), 4 Rue Enrico Fermi, F-69622 Villeurbanne Cedex, France}

\author{Navdeep Arya}
\affiliation{Department of Physics, Stockholm University, SE-106 91 Stockholm, Sweden}

\author{Simon Baier}
\affiliation{Universität Innsbruck, Institut für Experimentalphysik, Technikerstraße 25, 6020 Innsbruck, Austria}

% ...

\begin{abstract}
In this white paper, we highlight the importance of the ($1-10~{\rm Hz}$) frequency range for laboratory tests of the quantum nature of gravity using the quantum gravity-induced entanglement of masses (QGEM) protocol. QGEM requires matter-wave interferometers with masses ($m\sim10^{-15}-10^{-14}~{\rm kg}$), brought within separations ($d\sim30-50~\mu{\rm m}$), while maintaining spatial superpositions of ($1-20~\mu{\rm m}$) and coherence for ($\tau\sim 0.1 - 1~{\rm s}$). These requirements make low-frequency environmental noise a central experimental challenge and place QGEM in a regime closely related to the low-frequency goals of the Einstein Telescope (ET) and the Cosmic Explorer (CE). In particular, QGEM is sensitive to relative acceleration noise (RAN) and to gravity-gradient noise (GGN) generated by seismic and other environmental mass-density fluctuations.
For representative parameters $m=10^{-14}~{\rm kg}$, $\Delta x=10~\mu{\rm m}$, and $\tau=1~{\rm s}$, the differential acceleration-noise amplitude spectral density must be suppressed well below the $10^{-15}~{\rm m\,s^{-2}/\sqrt{Hz}}$ level to keep 
acceleration-induced dephasing below the relevant experimental scale.
Achieving this level of low-frequency noise suppression is therefore a key requirement for QGEM and closely parallels the seismic and gravity-gradient noise challenges that ET and CE address.
\end{abstract}

\maketitle


\section{introduction}

Matter-wave interferometers provide exquisitely sensitive probes of weak forces and fundamental interactions through measurements of relative quantum phases. For two interferometer paths, the phase difference is determined by the difference in action,
$
\Delta\phi=\nicefrac{\Delta S}{\hbar},
$
where \(\hbar\simeq10^{-34}~{\rm kg\,m^2\,s^{-1}}\). Consequently, even an extremely small action difference, \(\Delta S\sim\hbar\), can generate an observable \({\cal O}(1)\) phase. This remarkable sensitivity makes matter-wave interferometry particularly well suited to probing gravity, despite the weakness of the gravitational interaction, \(G=6.67\times10^{-11}~{\rm m^3\,kg^{-1}\,s^{-2}}\). The gravitationally induced phase due to the Earth's field was first measured using neutron interferometry~\cite{Colella:1975,Werner:1979gi}. Since then, remarkable experimental progress has extended matter-wave interferometry from atoms~\cite{Fixler:2007is,Asenbaum:2016djh,Schaff:2014jfw,Kovachi_2015,Margalit:2020qcy} to increasingly massive molecules and mesoscopic objects~\cite{Arndt:1999kyb,Eibenberger:2013cqb,Haslinger:2014rxx,Arndt:2015qcd,Fein:2019dgf,Pedalino:2026tur}. These advances now make it realistic to address significantly more challenging questions concerning the interplay between quantum mechanics and gravity.

Recently, it has been proposed that a matter-wave interferometer could be used to test the quantum nature of gravity~\cite{Bose:2017nin, ICTS,marletto2017gravitationally}. Arguably, understanding the nature of gravitational interaction with matter is one of the outstanding questions in modern physics. A prime question is whether gravity is a quantum entity subject to the rules of quantum mechanics, such as the principle of linear superposition.
 
If gravity is treated on an equal footing with matter as part of a quantum system, matter-wave interferometers provide a promising test bed for probing quantum aspects of gravity at low energies. Typically, experiments use matter degrees of freedom, and both massless spin-2 gravitons and matter are quantised~\cite{Toros:2020krn,Mazumdar:2026ewn}, yielding tripartite entanglement between the two helicity states of the graviton and matter in spatial superposition~\cite{Rufo:2024ulr}. 

For experimental purposes, where the background curvature is extremely weak, we can quantise both matter and massless excitations of the graviton around a Minkowski background.
The computations can be performed in the context of low-energy effective field theory; see~\cite{Gupta:1954zz,Gupta:1968zz,Barker:1966zz,Donoghue:1994dn}. For a review, see~\cite{Scadron:1991ep}. Entanglement is a purely quantum feature that exhibits quantum correlations in complementary conjugate bases, which is richer than classical correlations~\cite{Horodecki:2009zz}.

One particular facet that stands out is the extraction of a quantum correlation between two masses in adjacent matter-wave interferometers in a laboratory setting, where masses are entangled solely due to the quantum nature of gravity. The mediation of such an entanglement is due to the virtual excitation of the graviton exchange between the two masses of the matter-wave interferometers~\cite{Bose:2017nin,Marshman:2019sne};  see also~\cite{marletto2017gravitationally} and the experimental white paper~\cite{Bose:2025qns}. The experimental protocol is known as 
the quantum gravity-induced entanglement of masses (QGEM) protocol. For theoretical input, see~\cite{Marshman:2019sne,Bose:2022uxe,BeckeringVinckers:2023kco}, and also~\cite{
Belenchia:2018szb, Christodoulou:2022mkf, Carney:2021vvt,Danielson:2021egj,oppenheim2023covariant}. 

The upshot of these discussions is that, under certain assumptions, it is possible to discover a massless spin-2 graviton in a lab indirectly as a quantum mediator. These assumptions are: (1) standard quantum mechanics, (2) general relativity, which ensures causality, and (3) locality, which ensures local matter-graviton interaction. 

The witnessing of gravity's quantum nature from the QGEM protocol can be justified \cite{Bose:2017nin,Marshman:2019sne} using the local operation and classical communication (LOCC) theorem~\cite{Bennett_1996}, which states that the two quantum systems cannot be entangled via a classical mediator; a quantum mediator is required to entangle them. In the QGEM context, it is pertinent to note that the graviton propagator is equivalent to quantum communication. In fact, a quantum gravitational interaction with matter entails entanglement between the two matter-wave interferometers \cite{Marshman:2019sne,Bose:2022uxe,Danielson:2021egj,BeckeringVinckers:2023kco}. A further check of this QGEM experiment can be provided by the analogue of the light-bending experiment in gravity in an opto-mechanical setup~\cite{Biswas:2022qto,Carney:2021vvt,Dutta:2026pty}. The essence of massless spin-2 graviton is crucial to obtain the correct light deflection classically and in perturbative treatment of quantum gravity; see~\cite{Scadron:1991ep} for a static source of matter, and for a rotating source of matter~\cite{Biswas:2026pem}. Confirming the graviton hypothesis in a lab will require one further step: observing entanglement between matter and photons via massless graviton exchange~\cite{Bose:2017nin,Biswas:2022qto}. However, the challenges for this experiment are 
even greater than those of the QGEM. The latter operates at low energies (in a non-relativistic limit), where we wish to see the microscopic origin of the Newtonian potential and whether the gravitational interaction with matter is classical or quantum. The QGEM experiment is also extremely helpful for collapse models, where gravity becomes inherently classical; a null result for gravity-induced entanglement could open a new interpretation of gravity, such as ~\cite{Penrose96,Diosi1987,kafri2014classical,tilloy2016sourcing,oppenheim2023postquantum}.

However, like any other experiment, the QGEM experiment is prone to many sources of noise, in particular, quantum decoherence~\cite {zeh1970interpretation,
Zeh:1973wxr,joos1985emergence,hornberger2003collisional,Martinetz:2022tya,Romero-Isart:2011sdw,Romero-Isart:2011yun}; for a review, see \cite{Schlosshauer:2014pgr,bassireview}. It is envisaged that the QGEM protocol will utilise two neutral spin-embedded nanodiamonds to create a spatial superposition via the Stern-Gerlach setup, which exploits an inhomogeneous magnetic field; see~\cite{Bose:2017nin, Pedernales20_GM,Marshman:2021wyk}. Hence, the major sources of dephasing/decoherence arise from the electromagnetic background and fluctuations~\cite{Fragolino:2023agd,Schut:2023tce}, decoherence due to air molecules~\cite{vandeKamp:2020rqh,
Tilly:2021qef,Schut:2021svd,Rijavec:2020qxd}, and system decoherence due to surface defects, spin coherence~\cite{WoodPRA22_GM,Zhou:2025jki}, phonon induced decoherence~\cite{Henkel:2021wmj,Xiang:2024zol}, current/magnetic field fluctuations~\cite{Moorthy:2025fnu}, motional thermality and generic force noise~\cite{Braccini_mass}, and various systematics related to initialisation of the experiment~\cite{Schut:2023eux}.

One of the most important challenges for the QGEM experiment is relative acceleration noise (RAN). 
Here, RAN refers to stochastic acceleration of each interferometric mass relative to its associated apparatus, projected along the direction of the spatial superposition. In the noise decomposition used below, RAN describes the contribution linear in position, while gravity-gradient noise (GGN) describes the tidal contribution, quadratic in position, arising from motion and density fluctuations of nearby matter, as well as seismic and atmospheric disturbances.
These noise sources have been analysed in the context of QGEM in Refs.~\cite{Toros:2020dbf,Grossardt:2020def,Wu:2024tcr,Wu:2022rdv,Wu:2024bzd,Moorthy:2026sbk}; for a review of gravity-gradient noise, see Ref.~\cite{Harms:2015zma}. Acceleration noise is also closely related to the Humpty-Dumpty problem in Stern-Gerlach interferometry~\cite{Schwinger:1988,Englert:1988,Scully:1989,Wu:2024tcr}: external forces perturb the relative positions and momenta of the spatially separated wave packets, preventing their perfect recombination and thereby reducing the interferometric contrast~\cite{Margalit:2020qcy}. Even when one aims to test aspects of gravitational nonclassicality somewhat different than its ability to perform quantum communication~\cite{hanif2024testing,das2026test}, or look at implementations of QGEM not explicitly using Stern-Gerlach interferometry~ \cite{yi2026spatial,krisnanda2020observable}, the requirement of massive particles in interferometry or delocalized at a similar scale inevitably remain, and will also be subject to the same RAN and GGN requirements.  

The central observation of this white paper is that protecting a QGEM experiment from RAN over an interferometric timescale \(\tau\sim0.1 - 1~{\rm s}\) places stringent requirements on environmental noise in the \(1\!-\!10~{\rm Hz}\) band. For representative parameters $m=10^{-14}~{\rm kg}$ and
$\Delta x\sim{\cal O}(10)~\mu{\rm m}$, the differential
acceleration-noise amplitude spectral density must be suppressed below the
$10^{-15}~{\rm m\,s^{-2}/\sqrt{Hz}}$ level.
Remarkably, this brings QGEM into a low-frequency noise regime closely related to that targeted by the Einstein Telescope (ET)~\cite{Hild:2010id,Harms:2022jth}, and Cosmic Explorer (CE)~\cite{hall2021gravitational}. The Einstein Telescope (ET) is a proposed third-generation underground gravitational-wave observatory designed to extend its sensitivity down to frequencies of a few Hertz, whereas Cosmic Explorer (CE) is a proposed surface-based observatory targeting frequencies down to approximately $5~\text{Hz}$. Environmental disturbances, particularly seismic and gravity-gradient (Newtonian) noise, limit the low-frequency performance of both designs. Advanced seismic isolation, environmental monitoring, and noise-mitigation strategies are therefore central to their low-frequency research programmes. These requirements make ET and CE particularly relevant to QGEM, whose coherence time of ($0.1-1~{\rm s}$) places it in a closely related ($1-10~{\rm Hz}$) noise environment.

In this white paper, we express the QGEM RAN requirement as an equivalent strain noise and compare it directly with ET and CE's low-frequency strain-noise budget. The comparable sensitivity requirements suggest that technologies, noise-characterisation methods, and mitigation strategies developed for low-frequency gravitational-wave detectors could play an important role in enabling laboratory tests of the quantum nature of gravity with QGEM.


\begin{figure}[ht]
    \centering
    \includegraphics[width=\linewidth]{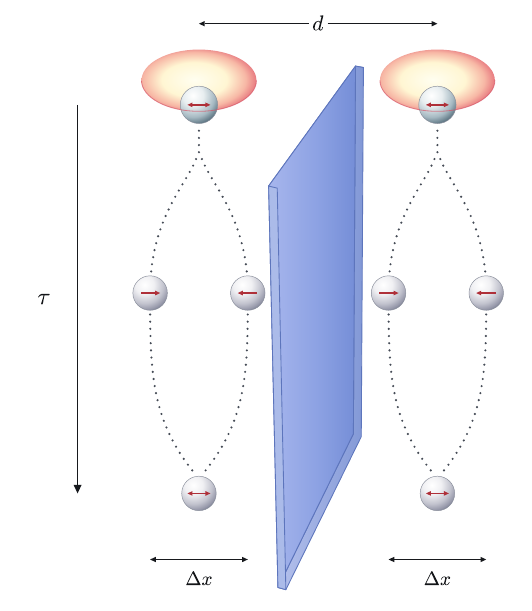}
    \caption{
    Schematic representation of the linear QGEM setup~\cite{Bose:2017nin}, in which two matter-wave interferometers are arranged as shown. A parallel configuration has also been proposed~\cite{Nguyen:2019huk,Tilly:2021qef}. In the linear setup, two nano-crystals with internal spins (red arrows) are initially trapped, motionally cooled, and separated by a thin membrane, ideally a superconducting shield, to suppress direct electromagnetic interactions between them. The shield is assumed to be sufficiently massive to be treated classically. Residual electromagnetic interactions between each nano-crystal and the shield have been studied in Refs.~\cite{vandeKamp:2020rqh,Schut:2023eux,Elahi:2024dbb}. Using standard radio-frequency techniques, each nano-crystal is prepared in a spin superposition,
$
\frac{1}{\sqrt{2}}\left(\ket{1}+\ket{-1}\right),
$
indicated by the double-sided arrows. After the traps are switched off, the particles pass through an inhomogeneous magnetic field, generating a Stern-Gerlach force~\cite{Bose:2017nin,Pedernales20_GM,Vicentini:2024noi,
Marshman:2021wyk}. The two spin components are accelerated in opposite directions, thereby creating a spatial superposition of size \(\Delta x\). 
During this stage, the gravitational interaction can entangle the two masses, each prepared in a spatial superposition.
A subsequent microwave spin flip reverses the acceleration, allowing the two branches to recombine. The final spin states are then measured using standard spin-readout techniques~\cite{Bose:2017nin,WoodPRA22_GM}. Repeating this sequence many times accumulates sufficient statistics to observe gravity-induced entanglement.}
    \label{QGEM_setup}
\end{figure}


\section{A brief introduction to the QGEM setup}

Based on the spatial configuration of the two interferometers, the anticipated setup of the QGEM protocol can be categorised into two setups: linear and parallel setups; see \cite{Nguyen:2019huk,Tilly:2021qef}.
The linear setup is schematically illustrated in Fig.~\ref{QGEM_setup}, where two superpositions are arranged in a linear configuration.
Spins are embedded in macroscopic masses (shown by arrows in balls). The parallel setup is similar to the linear setup, but the two superpositions are parallel; see details in \cite{Schut:2025blz}. 

We will witness the entanglement arising from the quantum nature of gravity via these two spin qubits embedded in a crystal, e.g., a micron-sized diamond crystal, which can host a nitrogen-vacancy (NV) center~\cite{WoodPRA22_GM} and has diamagnetic properties for levitation; see \cite{Elahi:2024dbb}. Spin is necessary in this context to create a quantum superposition via the Stern-Gerlach (SG) effect along with the readout
\cite{Bose:2017nin}; see experiments with atom interferometers in the SG setup~\cite{Margalit:2020qcy,Amit:2019cdk}. There are now many conceptual designs for creating massive spatial superpositions in the context of QGEM; see~\cite {Pedernales20_GM,Marshman:2021wyk,Zhou:2022jug,Zhou:2022frl,Zhou:2022epb}, where we imply individual SG interferometers. For the linear setup shown in Fig.~\ref{QGEM_setup}, the entanglement generated is a function of the ``entanglement phase" given by~\cite{Bose:2017nin,Marshman:2019sne}
\begin{equation} \label{Deltaphi}
    \Delta\phi_{\rm ent}=\frac{Gm_1m_2\tau}{\hbar }\left[ \frac{1}{d+\Delta x}+\frac{1}{d-\Delta x}-\frac{2}{d}\right]\,.
\end{equation}
In the limit $\Delta x\ll d $, we obtain
\begin{equation}
\Delta\phi_{\rm ent}\approx\left[\frac{2Gm_1m_2}{\hbar d}\right]\cdot\left[\frac{\Delta x}{d}\right]^2\tau,
\end{equation}
where $d$ is the distance between the centre of mass of the two adjacent interferometers, 
$\Delta x$ is the spatial superposition size, assumed to be the same for both interferometers, and we can assume $m_1=m_2=m$ as well, and $\tau$ is the time of the one-loop interferometer, assumed to be the same for both interferometers.
For simplicity, we approximate the branch separation as constant
at $\Delta x$ during the effective interaction time $\tau$, and
neglect the splitting and recombination stages. We use the same
approximation for the noise estimates below.
Note that the entanglement phase can be made order unity for certain choices of parameters. 
In the original proposal~\cite{Bose:2017nin}, we considered
masses of order $10^{-14}\,\mathrm{kg}$, superposition sizes
and minimum branch separations of a few hundred micrometres,
and interaction times of a few seconds. 
However, the constraints on these quantities have now been slightly relaxed by introducing electromagnetic shielding between the two spatial superpositions~\cite{vandeKamp:2020rqh,Schut:2023eux,Schut:2023hsy}, which has reduced the demand on the sizes of the spatial superpositions, $\Delta x\sim 1-10~{\rm \mu m}$ and $d_{min}\sim 30-50~{\rm \mu m}$, and on the entanglement witness, which we discuss below. It is to be noted that the entanglement phase and witness can also be computed using a full wavefunction treatment of the two gravitationally interacting Stern-Gerlach interferometers, and is found to give similar results and time requirements \cite{Braccini_mass,Braccini_superposition}. However, performing the experimental run for $\tau \sim 1$~s poses problems for RAN, seismic, and GGN, as well as for noise arising from the instrumentation, which we will discuss now.

The time over which coherence can be maintained for a state with a given superposition size depends on the dephasing rate $\gamma$ and on the value required for the entanglement witness. We envisage observing the entanglement by measuring NV spin-spin correlations between the two interferometers. Moreover, this time can be made smaller, albeit at the cost of
a smaller spatial superposition. It depends on the coherence time of the embedded spin in a nanoparticle. The spatial superposition experiment is limited in duration by the NV spin coherence time. Nanodiamond spin coherence times of approximately $780\,\mu\mathrm{s}$ have been reported at room temperature using spin-locking~\cite{March23_GM, WoodPRB22_GM, Frangeskou18_GM}. The coherence time is increasingly getting better for NV-centered systems. For instance, the NV electron-spin coherence time in bulk diamond can exceed $1~\text{s}$ at $3.7~\text{K}$ through dynamical-decoupling sequences tailored to the surrounding $^{13}\mathrm{C}$ nuclear-spin environment; see~\cite{Abobeih_2018}. At sub-kelvin temperatures, the coherence time may improve further.

The total dephasing rate can be summed up as
%
\begin{equation}
\gamma=\gamma_{\rm D}+\gamma_{\rm RAN}+\gamma_{\rm GGN}\,,
\end{equation}
where we discuss all these 
dephasing terms below.

\begin{enumerate}
\item $\gamma_{\rm D}$ denotes all the sources of decoherence which are intrinsic to the experimental setup, where the nanoparticle is levitated, motionally cooled and during the creation of the spatial superposition. 
This includes blackbody emission and absorption, internal heating,
and collisional decoherence~\cite{Bose:2017nin,vandeKamp:2020rqh,Schut:2024lgp}.
For example, for $m\sim10^{-15}$~kg and a superposition size
$\Delta x\sim10\,\mu\mathrm{m}$, an ambient temperature of
$4$~K and a pressure of order $10^{-13}$~Pa give a collisional
decoherence rate of order $0.1$~Hz; see~\cite{Schut:2024lgp}.
There are also electromagnetic interactions such as electric and magnetic dipole-dipole interactions with ambient gas molecules~\cite{Fragolino:2023agd}, Casimir interactions at leading order with the electromagnetic shielding plate~\cite{vandeKamp:2020rqh,
Schut:2023eux,Elahi:2024dbb}, magnetic flux from superconducting sources, current fluctuations, and intrinsic effects in the nanodiamond based on surface and interior impurities~\cite{Zhou:2025jki}. 
The decoherence rate $\gamma_{\mathrm{D}}$ also contains contributions from other electromagnetic fluctuations in the vicinity of the experiment~\cite{Fragolino:2023agd,Schut:2023tce}.
In summary, these sources can be categorised as noise in quantum spectroscopy.
Repeated runs provide the statistics needed to estimate
the entanglement witness and to characterize preparation
and readout errors~\cite{Schut:2023eux}.
These errors are treated separately from the physical
decoherence rate $\gamma_{\mathrm{D}}$, which includes contributions
from noise associated with material defects and impurities, levitation noise, and the
rotational degrees of freedom of the rigid body;
see~\cite{Tilly:2021qef,Schut:2021svd}.

\item In addition to quantum decoherence, the leading-order dephasing effect for the QGEM experiment comes from the relative acceleration noise (RAN), denoted by $\gamma_{\rm RAN}$, which we had discussed briefly. This kind of noise is prevalent due to any random acceleration, which will impact the trajectories of the spatial quantum superposition during the experiment. 

The RAN occurs at a linear order in position; e.g., see~\cite{Toros:2020dbf}. The corresponding interaction Lagrangian is
${\cal L}_j\sim m a(t)x_j$, where $a(t)$ is the stochastic acceleration relative to the apparatus and $j=1,2$ labels the two interferometer paths.

RAN could arise from any acceleration noise we cannot account for in the experiment~\cite{Toros:2020dbf,Moorthy:2026sbk}. RAN can be mitigated in an experimental design by suspending the experimental chamber, as with the ET mirrors~\cite{ET:2025xjr}. 

\item Higher-order effects arise in gravity gradient noise (GGN), also known as Newtonian Noise. 
GGN is a noise contribution of higher order in the spatial displacement than RAN~\cite{Saulson:1984,HughesThorne:1998}, e.g., the Lagrangian of the matter-wave interferometer, 
${\cal L}_j\sim m\omega^2_{\rm gg}(t)x_j^2,$ where $\omega_{\rm gg}$ is the time-varying stochastic quantity describing the \emph{gravity-gradient} potential, and $j=1,2$ are the two paths of the interferometer. GGN can be caused by ambient atmospheric conditions outside the experimental chamber, vehicle motion, seismic activity, and human activity in the vicinity of the experiment~\cite{ThorneWinstein:1998}. GGN can be mitigated by creating zones of isolation; see the detailed analysis for the QGEM experiment in  ~\cite{Toros:2020dbf,Grossardt:2020def}.

\end{enumerate}

In this paper, we focus mainly on RAN and discuss the overlap between the requirements for the QGEM experiment and those for the strain amplitude spectral density (ASD) of the ET and CE.

For the QGEM experiment, the entanglement witness, which is determined by the PPT (positive partial transpose) criterion~\cite{Horodecki:2009zz}, is given by~ \cite{Chevalier:2020uvv}; see also \cite{Schut:2021svd,Schut:2025blz}~\footnote{
Once the witness operator is fixed for the target state, its expectation value can be estimated from local spin
measurements and their correlations, without full state tomography. In the original paper \cite{Bose:2017nin}, we used a witness similar to Bell's inequality. However, it is now customary in the QGEM literature to use the entanglement witness $\mathcal{W}$, which is defined by the Positive Partial Transpose (PPT) witness~\cite{PeresPPT}. This criterion was first introduced for the QGEM experiment by ref.~\cite{Chevalier:2020uvv}, because it helps detect entanglement in bipartite systems cleanly as compared to the Bell type correlations.
The PPT witness is obtained by taking the partial transpose of the projector onto the normalized eigenvector corresponding to the most negative eigenvalue of the partially transposed
density matrix:
\begin{equation}
    \mathcal{W} = \left(\ket{\lambda_-^{T_2}}\bra{\lambda_-^{T_2}}\right)^{T_2} \, , \label{eq:ppt_def2}
\end{equation}
where $\lambda_-^{T_2}$ is the most negative eigenvalue of the partial transpose of the density matrix (denoted $\rho^{T_2}$). The expectation value of the PPT witness is given as:
\begin{equation}
    \langle {\cal W} \rangle 
    \equiv \Tr({\cal W}\rho) = \Tr(\ket{\lambda_-^{T_2}}\bra{\lambda_-^{T_2}}\rho^{T_2}) = \lambda_-^{T_2} \,. \label{eq:ppt_def}
\end{equation}
} \footnote{The formula Eq.~\eqref{Witness} is valid for both linear and parallel setups of the QGEM experiment; see~\cite{Schut:2025blz}. For the parallel setup, the entanglement phase should be replaced by $\Delta\phi_{\rm ent}=\frac{2Gm^2\tau}{\hbar}\left(\frac{1}{\sqrt{d^2+\Delta x^2}}-\frac{1}{d}\right)$; see~\cite{Schut:2025blz}. Therefore, the dephasing analysis below also applies to the parallel setup of the QGEM experiment. Hence, for fixed $m$, $\Delta x$, $\tau$, and dephasing budget, the RAN sensitivity $(S_{aa})_{\rm max}^{1/2}$ does not change, nor does the final expression for the strain noise $S_h^{1/2}$.}.
For this benchmark calculation, we assume equal, independent local dephasing in the two interferometers, with local dephasing factors \(e^{-\gamma\tau}\). Correlated dephasing, such as that arising from shared apparatus motion, requires evaluating the witness using the corresponding joint noise model.
To leading order for $\gamma\tau\ll1$ and $\lvert\Delta\phi_{\rm ent}\rvert\ll1$, the PPT-witness expectation value simplifies to
\begin{equation}\label{Witness}
\langle {\cal W}\rangle \simeq \frac{1}{4} \left( 2\gamma\tau -\vert\Delta \phi_{\rm ent}\vert\, \right).
\end{equation}
Here, $\gamma$ denotes the total dephasing/decoherence rate, which we need to minimise to protect the system's coherence; see the above discussion. For witnessing the entanglement between the two interferometers, the entanglement witness has to be negative~\cite{Chevalier:2020uvv,Schut:2021svd}, namely \begin{equation}
\langle {\cal W}\rangle<0.
\end{equation}
Hence, it requires that the entanglement phase be larger than twice the total dephasing $\Gamma=\gamma \tau$.

In principle, GGN is a higher-order noise contribution in the spatial displacement than RAN. For interferometric arms at the micron separation, such as QGEM, the dephasing caused by GGN is much smaller than that caused by RAN~\cite{Toros:2020dbf}. Assuming that RAN dominates the dephasing budget, we identify
$\gamma\tau\simeq\Gamma_{\rm RAN}/2$, where $\Gamma_{\rm RAN}$ denotes
the RAN-induced phase variance. The condition for witnessing entanglement
then becomes
\begin{equation} \label{eq:condition}
\lvert\Delta\phi_{\rm ent}\rvert>\Gamma_{\rm RAN}.
\end{equation}
In this white paper, we focus on RAN-induced dephasing and translate the corresponding acceleration-noise level into an equivalent strain noise for the ET parameters, with a baseline of $L=10$~km~\footnote{The CE baseline is $L=40$~km, and our equations can be normalised with respect to that of the CE also. However, for brevity, we show our results normalised with regard to that of ET. We will show the strain sensitivity for both ET and CE.}.

\section{Strain-noise equivalent for the QGEM experiment}
The QGEM protocol requires maintaining coherent spatial superpositions of mesoscopic masses over an experimental timescale \(\tau \sim 0.1 - 1\,\mathrm{s}\), corresponding to the low-frequency band \(f \sim 1\text{-}10\,\mathrm{Hz}\). In this regime, noise can degrade the negativity of the spin-entanglement witness, $\langle {\cal W} \rangle$, and hence conceal the delicate quantum-mechanical signature of gravity. Here we focus on relative acceleration noise, but a similar analysis can be carried out, for example, for gravity gradient noise, or other sources of noise~\cite{Toros:2020dbf}. 

The acceleration-induced phase diffusion may be estimated as~\cite{Toros:2020dbf}
\begin{equation}\label{RAN-0}
    \Gamma_{\mathrm{RAN}}\equiv \mathbb{E}\left[\Delta \phi^2\right]  \simeq \frac{S_{aa} \Delta x^2 m^2 \tau}{\hbar^2},
\end{equation}
where $\mathbb{E}[\,\cdot\,]$ denotes the expectation value over different noise realizations, $\Delta \phi$ corresponds to phase fluctuation induced by the relative acceleration noise, $S_{aa}$ denotes the acceleration noise power spectral density, $\Delta x$ is the superposition size (assumed equal for the two interferometers), $m$ is the mass of the two particles (for simplicity also assumed equal), and $\tau$ is the experimental timescale. 
Throughout, $S_{aa}(f)$ denotes the one-sided acceleration-noise power spectral density in Hertz. Equation~\eqref{RAN-0} is an order-of-magnitude white-noise estimate obtained using an effective bandwidth of order $1/\tau$; numerical factors of order unity depend on the detailed interferometer trajectory and the PSD convention.
Here, $\Gamma_{\mathrm{RAN}}$ denotes the RAN-induced phase
variance. For zero-mean Gaussian phase noise, the corresponding
dephasing factor is $\exp(-\Gamma_{\mathrm{RAN}}/2)$.
Note that the above expression Eq.~\eqref{RAN-0} was derived in the linear configuration of the QGEM setup shown in Fig.~(\ref{QGEM_setup}).  However, we have checked that this expression will not change in the parallel configuration either.


In the linear configuration, the centre-to-centre separation
$d$ and the minimum branch separation $d_{\min}$ are related
by $d=d_{\min}+\Delta x$. For the illustrative choice
$d_{\min}=30\,\mu\mathrm{m}$, superposition sizes
$\Delta x\simeq 1$--$10\,\mu\mathrm{m}$ correspond to
$d\simeq 31$--$40\,\mu\mathrm{m}$. For the benchmark used
below, we choose $\Delta x=10\,\mu\mathrm{m}$ and hence
$d=40\,\mu\mathrm{m}$.

The expression for $\Gamma_{\rm RAN}$, Eq.~(\ref{RAN-0}) becomes
\begin{widetext}
   \begin{equation}\label{Gamma_RAN}
       \Gamma_{\rm RAN}\approx 0.9 \left(\frac{m}{10^{-14}{\rm kg}}\right)^2\left(
       \frac{\Delta x}{10{\rm \mu m}}\right)^2
       \left(\frac{S_{aa}^{1/2}}{10^{-15}{\rm m\,s^{-2}/\sqrt{Hz}}}\right)^2\left(\frac{\tau}{1{\rm s}}\right)\,.     
   \end{equation} 
\end{widetext}
Eq.~\eqref{Gamma_RAN} expresses the accumulated RAN-induced phase variance in terms of the acceleration-noise ASD and the
interferometer parameters. To formulate the corresponding noise requirement, we introduce $\Gamma_{\max}$ as the
allocated upper bound on this phase variance. This allowance is chosen to satisfy the entanglement-witness condition for
the experimental parameters under consideration. Requiring $\Gamma_{\mathrm{RAN}}\leq\Gamma_{\max}$ and inverting
Eq.~\eqref{Gamma_RAN}, we obtain the following estimate of the maximum tolerable acceleration-noise ASD:
\begin{widetext}
   \begin{equation}
(S_{aa})^{1/2}
\approx 1.8\times 10^{-16}\left(\frac{10^{-14}{\rm kg}}{m}\right)\left(\frac{10{\rm \mu m}}{\Delta x}\right)\left(\frac{1{\rm s}}{\tau}\right)^{1/2}
\left(\frac{\Gamma_{\max}}{3.0\times 10^{-2}}\right)^{1/2} {\rm m s^{-2}/\sqrt{Hz}}\,.
\label{acc-noise}
\end{equation}
\end{widetext}
%
For the benchmark $m=10^{-14}\,\mathrm{kg}$,
$\Delta x=10\,\mu\mathrm{m}$, $d=40\,\mu\mathrm{m}$, and
$\tau=1\,\mathrm{s}$, Eq.~(1) gives
$|\Delta\phi_{\mathrm{ent}}|\simeq 0.2$.
Assuming $\gamma_{\mathrm{D}}$ and $\gamma_{\mathrm{GGN}}$ are negligible,
we choose $\Gamma_{\max}=3\times10^{-2}$ as a conservative
allowance for RAN-induced phase variance, satisfying
$\Gamma_{\max}<|\Delta\phi_{\mathrm{ent}}|$
as required by Eq.~\eqref{eq:condition}.
At $\Gamma_{\mathrm{RAN}}=\Gamma_{\max}$, Eq.~\eqref{Witness} gives
$\langle W\rangle\simeq -4.5\times10^{-2}$~\cite{Schut:2025blz}.
At fixed $\Gamma_{\max}$, increasing $\Delta x$ tightens the
acceleration-noise requirement, while shortening $\tau$
relaxes it. These scalings describe a fixed dephasing
allowance; changing either parameter also changes the
gravitational entangling phase. The entanglement condition
in Eq.~\eqref{eq:condition} must therefore be checked for each parameter choice.
We have kept the mass considered in the original proposal $m\sim 10^{-14}$kg.  This range of mass helps to make the witness negative. However, with meticulous design that creates a superposition with the electromagnetic shielding, we were able to reduce $\Delta x$ and $d_{min}$; see~\cite{vandeKamp:2020rqh,Schut:2021svd,Schut:2023hsy,Schut:2025blz}.  It is worthwhile to note that tabletop vibration-isolation systems can reach residual acceleration-noise amplitude spectral densities of order $(S_{aa})^{1/2}\sim 10^{-8}~{\rm m\,s^{-2}/\sqrt{Hz}}$ around $1~\text{Hz}$ with the help of active isolation and multi-stage suspensions~\cite{Kirchhoff_2020}. 
This suggests that for a lower mass $m\sim 10^{-18}$~kg, we can arrange a combination of $\Gamma,~\tau, \Delta x$ such that we can tolerate the acceleration noise of order $(S_{aa})^{1/2}\sim 10^{-8}~{\rm m\,s^{-2}/\sqrt{Hz}}$, when $\Delta x \sim 1 ~{\rm nm}$, 
$\tau\sim 1~\text{s}$
and $\Gamma\sim {\cal O}(10^{-2})$. For these parameters, the gravitational entangling phase would be too small to test the quantum nature of gravity. Nevertheless, experiments in this regime may be accessible to existing laboratory platforms and could serve as a useful stepping stone toward the larger masses, spatial separations, and coherence times required for QGEM.

Now, we can also convert this acceleration noise $S_{aa}^{1/2}$ into a strain noise using $x(f)=a(f)/(2\pi f)^2$, namely
\begin{equation} \label{eq:Sh}
S_{h}^{1/2}=\frac{S_{aa}^{1/2}}{(2\pi f)^2L}\, ,
\end{equation}
where \(S_h^{1/2}\) denotes the equivalent strain-noise amplitude spectral density and \(L\) is a reference baseline. The purpose of this conversion is to express the QGEM acceleration-noise requirement in the strain units used for gravitational-wave detectors. In the free-mass approximation, the same acceleration-noise ASD acting on a detector’s differential test-mass coordinate would produce the equivalent strain ASD given by Eq.~\eqref{eq:Sh}. We therefore use \(L=10\,\mathrm{km}\) for ET~\cite{ET:2025xjr} and \(L=40\,\mathrm{km}\) for CE~~\cite{hall2021gravitational}. These baselines enter only through the conversion and are not physical dimensions of the QGEM apparatus.

The comparison identifies frequency ranges in which the QGEM-equivalent strain requirement is comparable to the projected detector noise budgets. This motivates investigating whether technologies developed to suppress acceleration noise in ET and CE can be adapted to QGEM. A quantitative assessment requires the relevant acceleration-noise contributions, their spatial correlations, and the QGEM interferometer response. Combining Eqs.~\eqref{RAN-0} and \eqref{eq:Sh}, we obtain

\begin{widetext}
   \begin{equation}\label{Sh_Gamma_RAN}
       S_{h}^{1/2} \approx  \left(\frac{2.7\times 10^{-22}}{{\rm \sqrt{Hz}}}\right)\left( \frac{\Gamma_{\rm RAN}}{0.01}\right)^{\frac{1}{2}} \left(\frac{10^{-14}{\rm kg}}{m}\right)\left(
       \frac{10{\rm \mu m}}{\Delta x}\right) \left(\frac{1{\rm s}}{\tau}\right)^{\frac{1}{2}} \left(\frac{\rm 10 km}{L}\right) \left(\frac{1{\rm Hz}}{f}\right)^2
       \,.   
   \end{equation} 
\end{widetext}
Applying Eq. \eqref{Sh_Gamma_RAN} and taking the ET baseline $L=10$km, we plot the spectral density of the strain noise resulting from different values of the phase variance $\Gamma_{RAN}$ arising from the QGEM experiment; see the colour band in the left plot of Fig. \ref{fig:qgem_et_equivalent_strain}.  For comparison, we also plot the strain sensitivity curves of
ET-B~\cite{Abernathy2011ET} and the ET-D low-frequency interferometer (ET-D-LF)~\cite{Hild:2010id}.
The data resources for ET sensitivity curves are provided by the ET design team; see Refs. 
\cite{Hild:2010id, ETD_sensitivity_2018, ET_sensitivity_2023}. 
In the $1-10~{\rm Hz}$ range considered here, the ET strain-sensitivity
curve overlaps the QGEM-equivalent strain band corresponding to
$\Gamma_{max}\sim \Gamma_{\rm RAN}\sim{\cal O}(10^{-2})$. This indicates comparable noise
scales; the actual QGEM dephasing depends on the local acceleration-noise
spectrum and the interferometer filter function. 

In addition, we also plot the equivalent strain by taking the CE's baseline, $ L=40$ km; see the colour band in the right plot of Fig. \ref{fig:qgem_et_equivalent_strain}, and we compare QGEM's strain sensitivity with the strain sensitivity of CE in $1-10$~Hz band. The colour band corresponds to $10^{-4}\leq\Gamma_{\rm RAN}\leq10^{-1}$.
This comparison shows that the ET and CE noise budgets provide
useful benchmarks for the RAN requirements of the QGEM experiment.

In summary, this comparison motivates investigating whether noise-mitigation techniques developed for ET and CE can be adapted to QGEM experiments with interrogation times of \(0.1–1\,\mathrm{s}\). Of course, this assumes that we have very good control over $\gamma_{\rm D}$ and $\gamma_{\rm GGN}$.

\section{Conclusion}

In this white paper, we argued that the QGEM experiment could benefit from noise-control technologies developed for ET and CE to witness quantum entanglement between two matter-wave interferometers, arising from the quantum nature of gravity, in the $1-10$ Hz regime. The QGEM experiment requires roughly $ 0.1-1$ s to complete the one-loop interferometer, which is essential to build sufficient entanglement to be witnessed in the presence of decoherence and dephasing~\cite{Schut:2025blz}. 

There are many sources of noise; some are intrinsic and can be controlled by the local experimental design, and can be categorised by the dephasing/decoherence parameter $ \gamma_{\rm D}$. Any gravitational experiment suffers from the GGN, which are higher-order effects in the trajectories of the matter-wave interferometer, parameterised by the dephasing parameter $\gamma_{\rm GGN}$. In the context of QGEM, there are ways to mitigate them by isolating the experiment; see~\cite{Toros:2020dbf}. However, 
the leading-order noise not included in $\gamma_{\rm D}$ is described by the relative acceleration noise affecting the matter-wave interferometer. For representative QGEM parameters, maintaining
$\Gamma_{\rm RAN}\sim10^{-2}$ requires an acceleration-noise ASD of order
$(S_{aa})^{1/2}\sim10^{-16}~{\rm m\,s^{-2}/\sqrt{Hz}}$. This is a stringent constraint on the QGEM experiment, which requires a separate level of attention in research and development besides creating the spatial superposition of $\Delta x \sim {\cal O}(1-10){\rm \mu m}$ of a massive object of $m\sim {\cal O}(10^{-14})$ kg. 

\begin{widetext}

\begin{figure}
    \centering
    \includegraphics[width=0.45\linewidth]{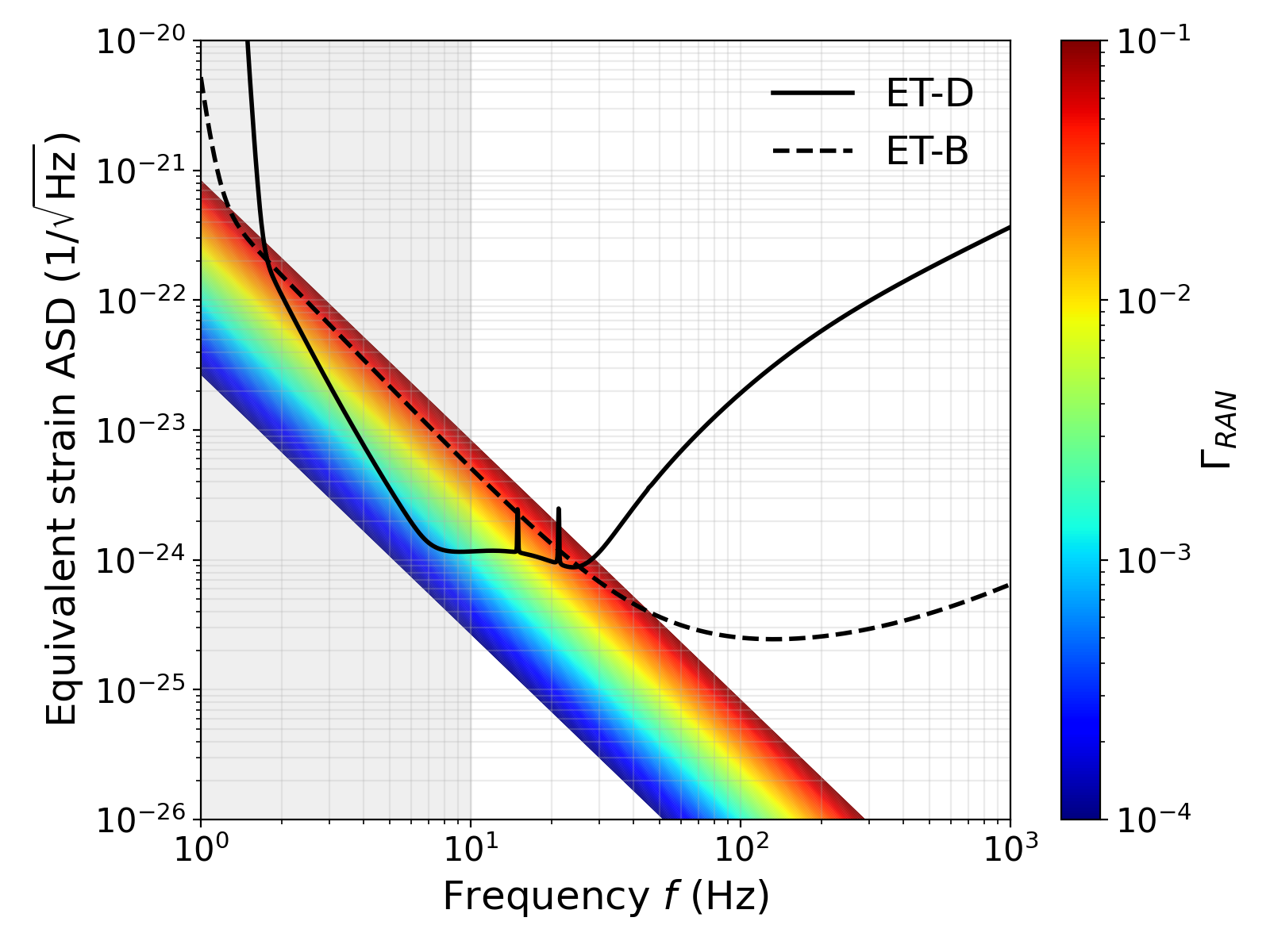}
    \includegraphics[width=0.45\linewidth]{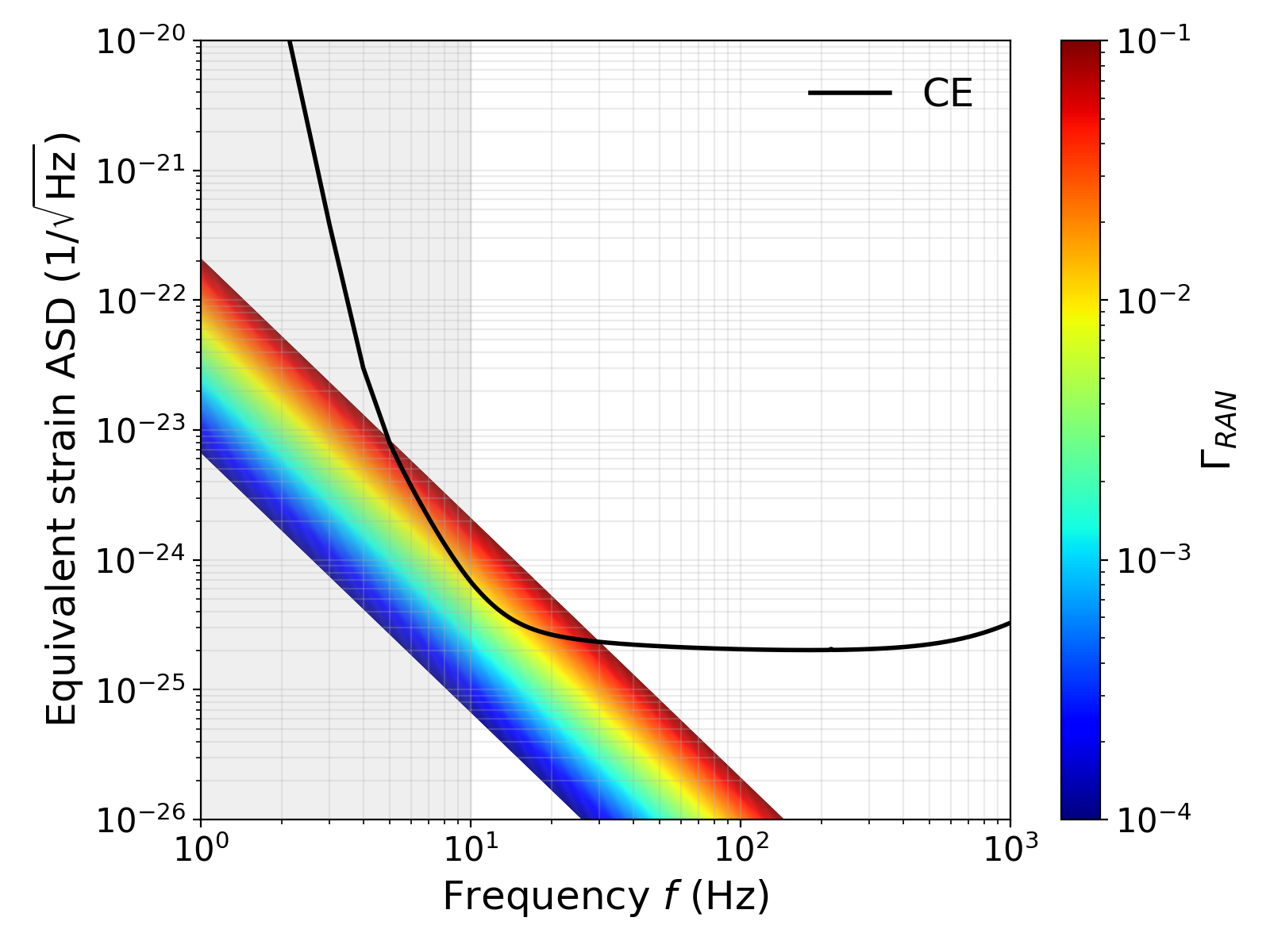}
    \caption{The left plot shows the comparison between the QGEM acceleration-noise requirement expressed as an ET-equivalent strain and the sensitivity of Einstein Telescope with $10$km baseline. The colour band shows the ET-equivalent strain ASD obtained from
Eq.~\eqref{Sh_Gamma_RAN} for dephasing values $10^{-4}\leq\Gamma_{\rm RAN}\leq10^{-1}$, taking $m=10^{-14}~{\rm kg}$, $\Delta x=10~\mu{\rm m}$, and $\tau=1~{\rm s}$. The solid black curve labelled ET-D shows the predicted strain-amplitude
spectral density of the ET-D low-frequency interferometer (ET-D-LF)~\cite{Hild:2010id}, while the dashed black curve shows
the ET-B strain-amplitude spectral density~\cite{Abernathy2011ET}. 
The vertical grey-shaded region indicates the $1-10~{\rm Hz}$ frequency range considered here, motivated by the characteristic QGEM timescales $\tau\sim0.1-1~{\rm s}$. The interferometer filter function determines the precise frequency dependence of the acceleration noise.
    In the right panel, the same acceleration-noise requirement is expressed
as a CE-equivalent strain ASD using $L=40~{\rm km}$, while the solid
curve shows the CE strain-noise budget~\cite{hall2021gravitational}. 
The overlap between the projected detector sensitivities and the QGEM-equivalent strain bands indicates comparable noise scales in parts of the low-frequency range. This motivates investigating whether noise-characterisation and mitigation techniques developed for ET and CE can be adapted to QGEM.
    }
    \label{fig:qgem_et_equivalent_strain}
\end{figure}
\end{widetext}

Fig.~\ref{fig:qgem_et_equivalent_strain} 
summarises the key results, showing the strain ASDs of ET and CE and their overlap with the equivalent strain-noise level required for the QGEM experiment to be performed to witness the quantum nature of gravity in a lab. The colour-coded regions depict $10^{-4} \leq \Gamma_{\rm RAN}\leq 10^{-1}$ in the range of $1-10$~Hz for both the gravitational wave observatories. 
Determining favourable QGEM interrogation times requires evaluating the accumulated dephasing using the local acceleration-noise spectrum and the interferometer filter function.
The different ET- and CE-equivalent curves reflect their respective
baselines and strain sensitivities and provide complementary benchmarks
for QGEM noise control.

For the parameters and chosen dephasing budget in Eq.~\eqref{acc-noise},  QGEM requires an acceleration-noise ASD of order \(10^{-16}\,\mathrm{m\,s^{-2}/\sqrt{Hz}}\). Expressing this requirement as an equivalent strain using the ET and CE baselines reveals comparable noise scales in parts of the low-frequency band. This comparison motivates investigating the adaptation of noise-characterisation and mitigation techniques from gravitational-wave observatories to QGEM.

\section{Acknowledgements} 
The first four authors conceptualised and wrote this white paper. All other authors carefully read the document, provided feedback, and support the paper's claims. 
A.M., G.W.M., A. G. and S.B.'s research is funded by the Gordon and Betty Moore Foundation through Grant GBMF12328, DOI 10.37807/GBMF12328. This material is based on work supported by the Alfred P. Sloan Foundation under Grant No. G-2023-21130. M.T.~acknowledges funding from the Slovenian Research and Innovation Agency (ARIS) under contracts N1-0392, P1-0416, SN-ZRD/22-27/0510 (RSUL Toro\v{s}). 
T. Z. is supported by the China Scholarship Council (CSC). 
N.~A. is supported by the Natural Sciences and Engineering Research Council of Canada (NSERC) and the University of Waterloo. Research at Perimeter Institute is supported in part by the Government of Canada through the Department of Innovation, Science and Economic Development Canada and by the Province of Ontario through the Ministry of Colleges and Universities.
A.S.A. is supported by Lembaga Pengelola Dana Pendidikan (LPDP) of the Ministry of Finance of Indonesia. 
N. A. is supported by Knut and Alice Wallenberg Foundation through a Wallenberg Academy Fellowship No. 2021.0119.
S. B. acknowledges funding by the Austrian Science Fund [10.55776/ESP258].
P. F. B. is supported by UKRI STFC and EPSRC  Grant Reference No(s):  ST/W006170/1, EP/W029626/1, EP/N031105/1. 
A. B. is supported by the University of Trieste and INFN. 
G. K. B. acknowledges support from the research programme N1-0392 and P1-0416 of the Slovenian Research and Innovation Agency (ARIS). 
L.B. would like to acknowledge Engineering and Physical Sciences Research Council (EPSRC) grants (EP/R513143/1 and EP/W524335/1).
The research of C.~C. was funded in part by the Austrian Science Fund (FWF) [10.55776/F71] and [10.55776/COE1].
M.~C. is supported by the University of Manchester.
L.~C. is supported by ESPRIT fellowship funded by Austrian Science Fund (FWF).
Y.~C. is supported by the Simons Foundation (Award No. 568762), the Heising-Simons Foundation, and the Keck Foundation. 
R. J. C. is supported by the Engineering and Physical Sciences Research Council EP/Y035046/1. 
A. D. C-D. acknowledges support from Project SA097P24 funded by Junta de Castilla y Le\'on (Spain), PID2024-158938NBI00 and CNS2024-154286 funded by MCIN/AEI/10.13039/\- 501100011033 and by {\it ERDF A way of making Europe}, and  NRF Grant CSUR23042798041 (South Africa). 
C.~C. is supported by INFN and by the European Union's Horizon Europe research and innovation programme under the Marie Skłodowska-Curie grant agreement No. 101236929 (NEXUS). 
S. D. acknowledges support by the European Research Council (ERC) under the European Union’s Horizon Europe research and innovation programme (ERC-2024-STG, 101165179, Art- DisQ) and from the German Research Foundation DFG (EXC 2064/1, Project 390727645).
S.~D. is supported by NSERC, Quantum Horizons Alberta. 
P. D. warmly acknowledges support from an NSERC discovery grant (RGPIN-2025-06182).
G. D. is supported by Grant No. 63121 from the John Templeton Foundation and Grant No. G-2023-21131 from the Alfred Sloan Foundation.
S. F. is supported by a studentship from the Engineering and Physical Sciences Research Council, grant number 2882481.
D. R. acknowledges funding from EPSRC (Engineering \& Physical Sciences Research Council, United Kingdom) Grant Number EP/X009467/1.
S. G. E. is supported by CIERA Board of Visitors Graduate Fellowship. 
F. F. acknowledges support from the Istituto Nazionale di Fisica Nucleare (INFN) through the Commissione Scientifica Nazionale 4 (CSN4) Iniziativa Specifica “Quantum Universe” (QGSKY). 
J. F. is supported by Japan Science and Technology through an EXPERT-J grant.
L. F. is supported by the Government of Canada through the Department of Innovation, Science and Economic Development and by the Province of Ontario through the Ministry of Colleges and Universities.
C. G. is supported through the action France 2030 from Agence Nationale de la Recherche, project OQULUS (ANR22-PETQ-0013).
M. G. is supported by the Einstein Foundation Berlin through an Independent Researcher Grant.
P.~T.~G. is supported by the project CZ.02.01.01/00/22\_010/0013054 (C-MONS) within Programme JAC MSCA Fellowships at Palacký University Olomouc IV.
M. G. is supported by EU Quantum Chips Act - DIREQT. 
A. G. acknowledges support from NSF grants PHY-2409472 and PHY-2110524, DARPA, the John Templeton Foundation, and the Simons Foundation. 
B. H. is supported by the European Union (ERC StG, CLOSEtoQG, Project 101041115) and the Dutch Research Council (NWO, SUMMIT.1.016 Quantum Limits). 
R.~H. acknowledges John Templeton Foundation grant 63683, as part of the WOST project.
D. H. is supported by National Academy of Sciences, India.
M.T.J. is supported by the Science and Technology Facilities Council on Quantum Simulators for Fundamental Physics (ST/T006900/1) as part of the UKRI Quantum Technologies for Fundamental Physics program. M.~T.~J. is supported by the Leverhulme Research Leadership Award (RL2019-020).
E.~K.-R. is supported by an EPSRC Doctoral Prize Fellowship (Grant No.~EP/W524335/1).
A. K. acknowledges financial support from the UK--Canada Quantum for Science research collaboration grant OPP640. 
A. K. was supported in part by Perimeter Institute for Theoretical Physics.
M. S. K. is supported by KIST. 
I. P. L. was partially supported by the National Council for Scientific and Technological Development - CNPq, grant 312547/2023-4.
A. L. is supported by NSERC through a Quantum Alliance grant. 
A. G. M. is supported by the “Table-top experiments for fundamental physics” program, sponsored by the Gordon and Betty Moore Foundation (Grant GBMF12328), Simons Foundation, Alfred P. Sloan Foundation (Grant No. G-2023-21130), and John Templeton Foundation.
M.~M.~M. is supported by the Austrian Science Fund (FWF) [10.55776/COE1].
C. M. is supported by The Gordon and Betty Moore Foundation.
F. M. is supported by CNRS (umr7588).
D.K.L.O. acknowledges funding from EPSRC (EP/T001011/1, EP/Z533208/1, EP/W027011/1), ESA (4000148330/25/NL/FGL/lf, 4000145824/24/NL/FFi), and EU COST Action RQI (CA23115).
M. G. A. P. is supported by the CSN5-T4QC initiative of INFN. 
M.P. is supported by the EU Horizon Europe EIC Pathfinder project QuCoM (GA no. 10032223). 
J. P. acknowledges the Knut and Alice Wallenberg Foundation through a Wallenberg Academy Fellowship No. 2021.0119.
T. K. P. is supported by the STFC under Grant No. ST/X003167/1. 
R.~R. is supported by Beasiswa Indonesia Bangkit and Lembaga Pengelola Dana Pendidikan (BIB LPDP) of the Ministry of Religious
Affairs and the Ministry of Finance of Indonesia.
A.~C.~R. is supported by a Ph.D. scholarship from the University of Salerno.
S.~Q. is funded by the Wallenberg Initiative on Networks
and Quantum Information (WINQ) at Nordita, where
Nordita is partially supported by NordForsk.
M. R. acknowledges support by the Wellcome Trust Early-Career Award 331016/Z/25/Z. 
S. M. is supported by the Czech Science Foundation through the JUNIOR STAR grant 25-17250M.
R. K. acknowledges support by the Slovenian Research and Innovation Agency (ARIS) under contracts no. P1-0416, N1-0501 and through the UL VIP project (KTTK21) under contract no. SN-ZRD/22-27/510. 
C. S. acknowledges financial support from AEI through Consolidaci\'on Investigadora (CNS2024-154149). 
A.~P. is partially supported by INFN grant FLAG.
B. C. S. is  supported by NSERC. M. S. is supported by the National Research Foundation, Singapore through the National Quantum Office, hosted in A*STAR, under its Centre for Quantum Technologies Funding Initiative (S24Q2d0009). 
M. E. T. is supported by the Australian Research Council grant number CE200100008.
G.~T. is supported by the MCIU (Grant No.~PID2025-171880NB-I00) and COST Action RQI (Grant No.~CA23115).
H. U. acknowledges funding from the EU Horizon Europe EIC Pathfinder project QuCoM (10032223), from the UK funding agency EPSRC (grants  EP/V035975/1, EP/V000624/1, EP/W007444/1, EP/X009491/1), and from the Leverhulme Trust (RPG-2022-57). 
G.~V. is supported by CNPq.
V. V. is supported by CNRS and ENS Paris.
F.~V. acknowledges support from the WOST (WithOut SpaceTime) project, funded by the John Templeton Foundation (Grant No.~63683).
The opinions expressed in this work are those of the authors and do not necessarily reflect the views of the John Templeton
Foundation.
V. V. is supported by the Gordon and Betty Moore Foundation and the Conjecture Institute. 
M.~V. is supported by the Ministry of Science, Technological Development and Innovations of the Republic of Serbia (MNTRI).
G.~F.~V. is supported by the University of Salerno.
C.~W. is supported by the Armenian Higher Education and Science Committee (Grant No.~24RL-1C024, Projects 21AG-1C024 and 25PostDoc1C003).
M. Z. is supported by WAF no. 2021.0119.
This white paper benefited from scientific exchange within COST Action CA23115, “Relativistic Quantum Information,” supported by COST (European Cooperation in Science and Technology).

\bibliography{Ref}

\end{document}